\documentclass[twocolumn,aps,prl,showpacs,preprintnumbers,groupedaddress]{revtex4}
\usepackage{amsmath}
\usepackage{graphicx}
\usepackage{bm}
\usepackage{color,colordvi}
\usepackage{dcolumn}
\begin{document}

\newcommand{\as}{\alpha_{\textrm s}}

\def\plus{{\!+\!}}
\def\minus{{\!-\!}}
\def\z#1{{\zeta_{#1}}}
\def\zz#1{{\zeta_{#1}^2}}

\def\ca{{C^{}_A}}
\def\cf{{C^{}_F}}
\def\tr{{T^{}_{\! R}}}

\def\caca{{C^{2}_A}}
\def\cfcf{{C^{2}_F}}

\def\nf{{n^{}_{\! f}}}
\def\n2f{{n^{\,2}_{\! f}}}

\def\nc{{N^{}_{\! c}}}
\def\n2c{{N^{\,2}_{\! c}}}

\def\lnr{{\,\ln\!\frac{M^2_\Phi}{\mu_r^2}}}
\def\lnf{{\,\ln\!\frac{M^2_\Phi}{\mu_f^2}}}
\def\lnrr{{\,\ln^2\!\frac{M^2_\Phi}{\mu_r^2}}}
\def\lnff{{\,\ln^2\!\frac{M^2_\Phi}{\mu_f^2}}}
\def\lnrf{{\,\ln\!\frac{\mu_r^2}{\mu_f^2}}}

\def\li#1{{\textrm{Li}_{#1}}}
\def\h{{\textrm{H}}}

\preprint{YITP-SB-04-37}
\preprint{BNL-HET-04/9}
\preprint{hep-ph/0407254}
\title{Higgs Boson Resummation via Bottom-Quark Fusion}
\author{B. Field}
\email[]{bfield@ic.sunysb.edu}
\affiliation{C.N. Yang Institute for Theoretical Physics, 
             Stony Brook University,
             Stony Brook, New York 11794-3840, USA}
\affiliation{Department of Physics, Brookhaven National Laboratory,
             Upton, New York 11973, USA}
\date{July 21, 2004}

\begin{abstract}

The region of small transverse momentum in $q\bar{q}-$ and
$gg-$initiated processes must be studied in the framework of resummation
to account for the large, logarithmically-enhanced contributions to
physical observables. In this letter, we study resummed differential
cross-sections for Higgs production via bottom-quark fusion. We find
that the differential distribution peaks at approximately $15$~GeV, a
number of great experimental importance to measuring this production
channel.

\end{abstract}

\pacs{13.85.-t, 14.80.Bn, 14.80.Cp}
\maketitle

Resummation of total and differential cross-section for the inclusive
production of a Higgs boson has concentrated on the gluon-gluon initial
state\cite{Catani:ne, Kauffman:1991jt, Yuan:1991we, Kauffman:cx,
Catani:1996yz, Kramer:1996iq, Balazs:2000wv, deFlorian:2000pr,
deFlorian:2001zd, Glosser:2002gm, Berger:2002ut, Berger:2003pd,
Bozzi:2003jy, Field:2004tt}. In the Standard Model (\textsc{sm}), the
gluon-gluon initial state gives the largest contribution to the total
and differential cross-sections, but this is not always the case in
extensions of the \textsc{sm}. In the Minimal Supersymmetric Standard
Model (\textsc{mssm}) the bottom-quark fusion initial state can be
greatly enhanced, perhaps leading to the observation of a supersymmetric
signal in nature, if the location of the peak of the differential
distribution in known.

The \textsc{mssm} contains two Higgs doublets, one giving mass to
up-type quarks and the other to down-type quarks. The associated vacuum
expectation values (\textsc{vev}s) are labeled $v_u$ and $v_d$
respectively, and fix the free \textsc{mssm} parameter $\tan\beta\equiv
v_u/v_d$. In the \textsc{mssm}, there are five physical Higgs boson mass
eigenstates. In this letter, we are interested in the neutral Higgs
bosons $\{ h^0, H^0, A^0 \}$ which we will call $\Phi$ generically.

In contrast to the \textsc{sm}, the bottom-quark Yukawa couplings in the
\textsc{mssm} can be enhanced with respect to the top-quark Yukawa
coupling. In the \textsc{sm}, the ratio of the $t\bar{t}\Phi$ and
$b\bar{b}\Phi$ couplings is given at tree-level by
$\lambda^{\textsc{sm}}_{t}/\lambda^{\textsc{sm}}_{b} = m_t/m_b \approx
35$. In the \textsc{mssm}, the coupling depends on the value of
$\tan\beta$. At leading order,

\begin{equation}
\label{eq::Yukrat}
\frac{\lambda^{\textsc{mssm}}_{t}}{\lambda^{\textsc{mssm}}_{b}} =
f_\Phi \, \frac{1}{\tan\beta}\cdot \frac{m_t}{m_b}\,,
\end{equation}
with
\begin{equation}
f_\Phi = \left\{
  \begin{array}{ll}
             - \cot\alpha\,,& \qquad \Phi=h^0 \, \\
    \phantom{-}\tan\alpha\,,& \qquad \Phi=H^0 \, \\
    \phantom{-}\cot\beta \,,& \qquad \Phi=A^0 \,
  \end{array}
  \right.
\end{equation}
where $\alpha$ is the mixing angle between the weak and the mass
eigenstates of the neutral scalars. Given the mass of the pseudoscalar
$M_{A^{0}}$ and $\tan\beta$, the angle $\alpha$ can be determined given
reasonable assumptions for the masses of the other supersymmetric
particles in the spectrum.

The form of $f_\Phi$ shows us that the production of the pseudoscalar
due to bottom-quark fusion is enhanced by a factor of $\tan^2\!\beta$,
which is a free parameter in the theory.

\section{The Bottom-Quark}
\label{bottom}

It is also important to define what is meant by a bottom-quark
distribution\cite{Olness:1987ep, Barnett:1987jw, Olness:1997yc}. In our
analysis, we employed the CTEQ6.1M bottom-quark parton
distribution\cite{Pumplin:2002vw} with $\as(M_Z) = 0.118$ and set the
mass of the Higgs boson of interest $M_\Phi = 120$~GeV. We compared the
bottom-quark distribution function in the PDF set and the numeric
solution to the Dokshitzer-Gribov-Lipatov-Altarelli-Parisi
(\textsc{dglap}) equations for a single quark splitting. A bottom-quark
distribution for a gluon splitting into a $b\bar{b}$ pair can be written
in the \textsc{dglap} formalism as

\begin{equation}
\tilde{b}(x,\mu) = \frac{\as(\mu)}{2\pi} 
                   \ln \biggl( \frac{\mu^2}{m_b^2} \biggr)
                   \int_x^1 \frac{dy}{y} 
                   P_{q \leftarrow g} 
                   \biggl( \frac{x}{y} \biggr) g(y,\mu),
\end{equation}
where $g$ is the gluon distribution, and the \textsc{dglap} splitting 
function is
\begin{equation}
P_{q \leftarrow g}(z) = \frac{1}{2} [ z^2 + (1-z)^2 ].
\end{equation}

The bottom-quark distribution is encoded into the CTEQ PDF set in this
manner\cite{fred}, but takes into account multiple quark splitting
functions. When evaluated with the \textsc{dglap} formalism, we found
the differential cross-section increased by approximately $10$\% near
its peak. However, we used the native bottom-quark distributions for
speed and to understand their built-in uncertainties.

Previously\cite{Field:2004tt}, we calculated in detail the resummation
coefficients for a differential cross-section for the scalar and
pseudoscalar Higgs boson from the gluon-gluon initial state. In this
letter, we will calculate the resummation coefficients needed for the
resummation of the $b\bar{b}$ initial state for the scalar and
pseudoscalar Higgs bosons in the same manner as the gluon-gluon channel
in Ref.~\cite{Field:2004tt}. We will leave the bottom-quark--Higgs
coupling set equal to the \textsc{sm} value so that the reader can scale
the results to whatever coupling value is of interest. 

\section{Resummation}
\label{resum}

The resummation formalism needs the lowest order total cross-section as
a normalization factor (see \cite{Field:2004tt} for details), $b\bar{b}
\rightarrow \Phi$ in this case. Following Ref.~\cite{Harlander:2003ai},
we will ignore the bottom-quark mass except in the Yukawa coupling with
the Higgs boson. Although the pseudoscalar Higgs couples to quarks with
a $\gamma_5$, there are no differences in the matrix elements (modulo
the \textsc{mssm} coupling factor) when the bottom-quark mass is
neglected.

It is important to use the $\overline{\textrm{MS}}$ running mass for the
bottom-quark in our calculation as the difference from the pole mass at
the scales involved is considerable\cite{Dicus:1998hs, Campbell:2002zm}.
In the \textsc{sm}, the bottom-quark Yukawa coupling is
$\lambda_b^{\textsc{sm}} = \sqrt{2} \overline{m}_b / v$, where $v$ is
the \textsc{sm vev} and is approximately equal to $246$~GeV and
$\overline{m}_b$ is the $\overline{\textrm{MS}}$ running mass. We have
set the bottom-quark mass $\overline{m}_b(\overline{m}_b) = 4.62$~GeV in
our calculations. The NLO running of the bottom-quark mass corresponds
to $\overline{m}_b(M_\Phi) = 3.23$~GeV. The coupling in the
\textsc{mssm} can be written

\begin{equation}
\begin{split}
\lambda_b^{\textsc{mssm}} &= \left\{
  \begin{array}{ll}
               \displaystyle
             - \sqrt{2} \, \frac{\overline{m}_b}{v}
                           \frac{\sin\alpha}{\cos\beta}
               \,,& \qquad \Phi=h^0 \, \\[15pt]
               \displaystyle
    \phantom{-}\sqrt{2} \, \frac{\overline{m}_b}{v}
                           \frac{\cos\alpha}{\cos\beta}
               \,,& \qquad \Phi=H^0 \, \\[15pt]
               \displaystyle
    \phantom{-}\sqrt{2} \, \frac{\overline{m}_b}{v}
                           \tan\beta
               \,,& \qquad \Phi=A^0. \,
  \end{array}
  \right.
\end{split}
\end{equation}

\begin{figure}
  \begin{center}
    \begin{tabular}{ccc}
      \resizebox{25mm}{!}{\includegraphics{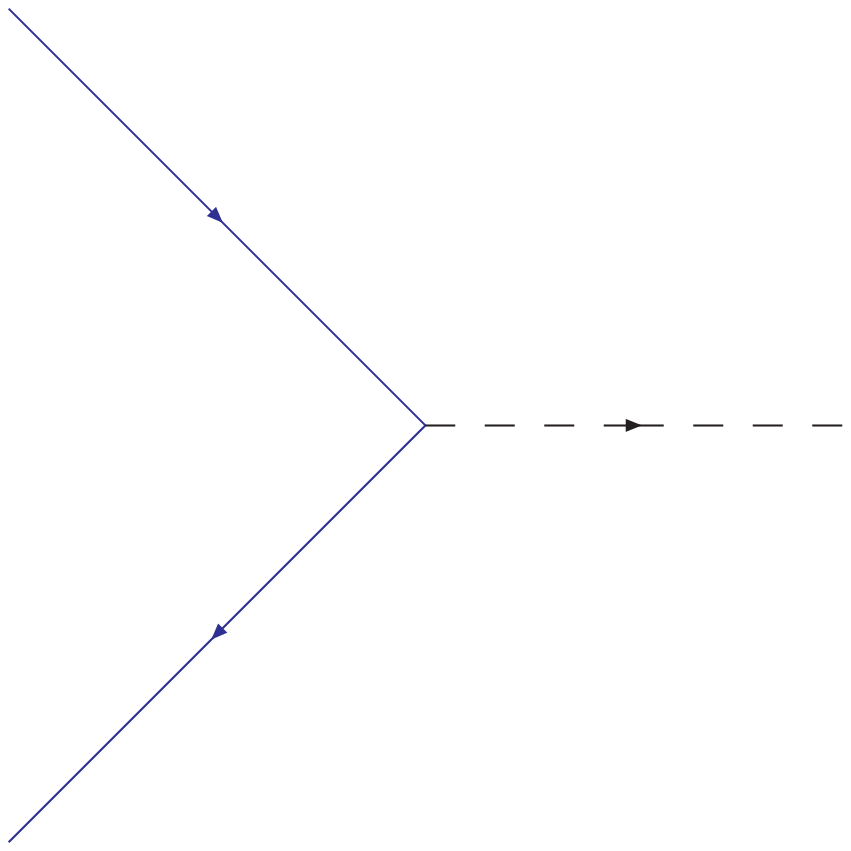}} &
      \resizebox{25mm}{!}{\includegraphics{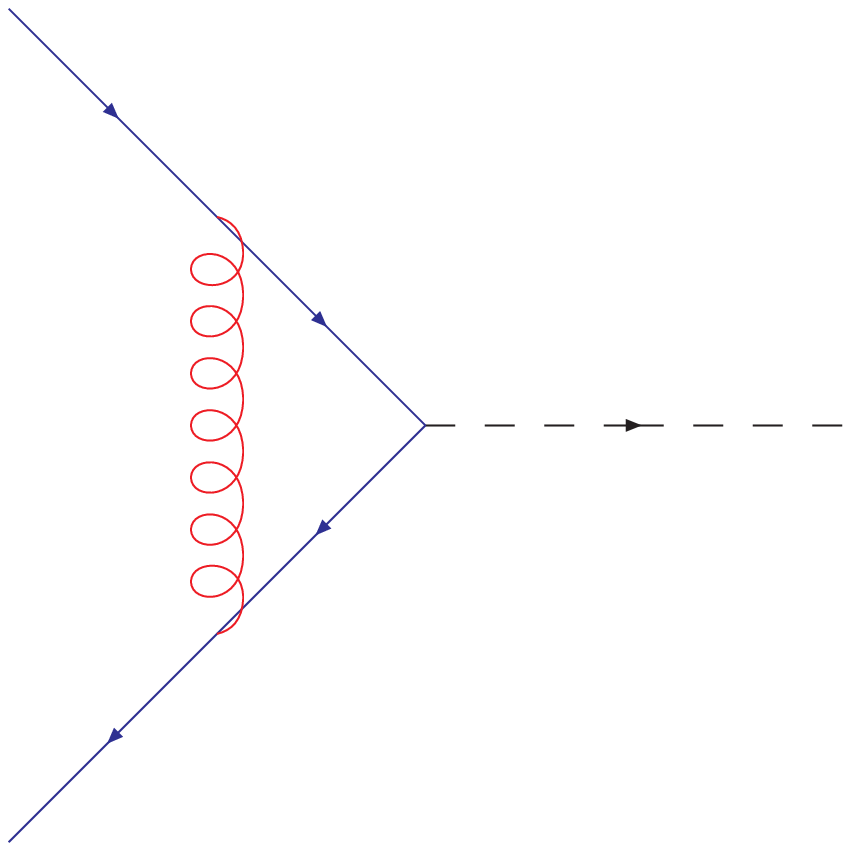}} &
      \resizebox{25mm}{!}{\includegraphics{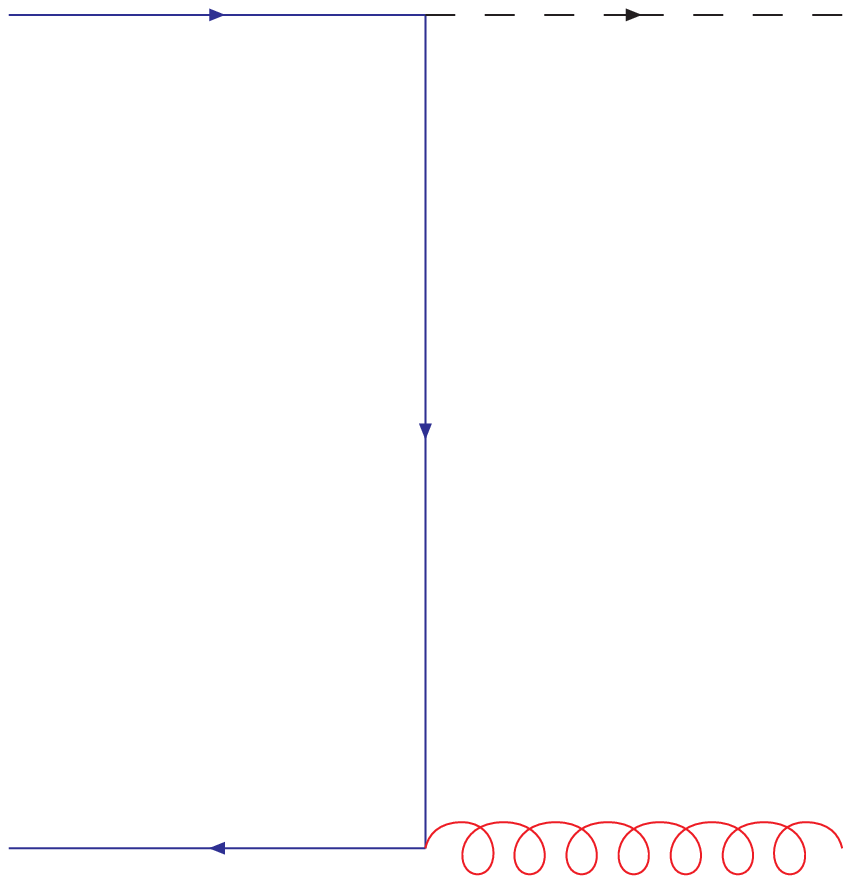}} \\
      (a) &
      (b) &
      (c)
    \end{tabular}
    \caption{Diagrams needed for the $b\bar{b}$ initial state resummed 
             differential cross-section. Figure~\ref{diags}a is the 
             lowest order production channel and couples differently for 
             different Higgs bosons. Figure~\ref{diags}b is the virtual 
             correction to the lowest order process. Figure~\ref{diags}c 
             is the lowest order graph contributing to the differential 
             cross-section. The crossed graph is not shown.}
    \label{diags}
  \end{center}
\end{figure}

The spin- and color-averaged total partonic cross-section (see
Fig.~\ref{diags}a) for the leading order subprocess, $b\bar{b}
\rightarrow \Phi$, can be easily written

\begin{equation}
\hat{\sigma}_0^{\textsc{sm}} 
               = \frac{6\pi}{4N_c^2}\frac{\overline{m}_b^2}{v^2}
                 \frac{1}{M_\Phi^2}\delta(1-z), 
\end{equation}
where $z = M_\Phi^2/\hat{s}$ and the number of colors $N_c = 3$. We also
need the LO differential cross-section (Fig.~\ref{diags}c) for the
next-to-leading log (NLL) resummation coefficients for the differential
cross-section. If we remove the $\delta(1-z)$ factor from our prefactor
$\hat{\sigma}_0$, then we can write the spin- and color-averaged
differential cross-section for $b(p_1)\bar{b}(p_2) \rightarrow g(-p_3) 
\Phi(-p_5)$ as

\begin{align} \nonumber
\frac{d\hat{\sigma}}{d\hat{t}} & = \hat{\sigma}_0 \biggl( \frac{\as}{\pi} 
                                                  \biggr)
         \frac{\cf}{2} \biggl(
         \frac{M_\Phi^4+\hat{s}^2}{\hat{s}\hat{t}\hat{u}}
         \biggr) \\
     & = \hat{\sigma}_0 \biggl( \frac{\as}{\pi}
         \biggr)
         \frac{\cf}{2}
         \frac{1}{p_t^2} \biggl[ 1 + z^2 \biggr].
\end{align}
where $\cf = (N_c^2-1)/2N_c$, and the kinematic variables are defined as
$\hat{s} = (p_1+p_2)^2$, $\hat{t} = (p_1+p_5)^2$, $\hat{u} =
(p_2+p_5)^2$, and $M_\Phi^2 = p_5^2$. In our second line, we have
written the differential cross-section in terms of $\hat{u}\hat{t} =
p_t^2 \hat{s}$ for the $2\rightarrow 2$ process. 

To find the resummation coefficients for a differential
cross-sections\cite{Kauffman:1991jt, Kauffman:cx, deFlorian:2001zd,
Field:2004tt} we integrate the differential cross-section around $p_t =
0$

\begin{equation}
\Delta\hat{\sigma} = \int_0^{q_t^2}
 dp_t^2 \frac{d\hat{\sigma}}{dp_t^2}
\end{equation}
and label this result `real' as it is similar to the real corrections to
the LO total cross-section. Working in $N=4-2\epsilon$ dimensions, we
find

\begin{align} \nonumber
\Delta\hat{\sigma}^{\textrm{real}} = \hat{\sigma}_0 z
\frac{\as}{\pi}
 \biggl[&\frac{\cf}{\epsilon^2} + \frac{3}{2}\frac{\cf}{\epsilon}
        -\frac{\cf}{2} \ln^2 \biggl( \frac{M_\Phi^2}{q_t^2} \biggr) \\
       +&\frac{3}{2} \cf \ln \biggl( \frac{M_\Phi^2}{q_t^2} \biggr)
        + \cf - \cf\z2
 \biggr].
\end{align}

To regularize this result, we need to add the virtual corrections that
are shown in Fig.~\ref{diags}b. These corrections are very similar to
Drell-Yan corrections\cite{Altarelli:1979ub}. The virtual corrections
can be written as

\begin{equation}
\Delta\hat{\sigma}_{\textrm{virt}} = \hat{\sigma}_0
     \biggl( \frac{\as}{\pi} \biggr)
     \biggl[ - \frac{\cf}{\epsilon^2} - \frac{3}{2}\frac{\cf}{\epsilon}
             - \cf + 2 \cf \z2 \biggr].
\end{equation}
In the Drell-Yan case, the $-\cf$ factor would be $-4\cf$. When the two
results are added together the resummation coefficients are easily read
off from the expression. The total expression is

\begin{align} \nonumber
\Delta\hat{\sigma} = \hat{\sigma}_0 z \biggl[
 1&+ \frac{\as}{\pi} \biggl(
    -\frac{\cf}{2} \ln^2 \biggl( \frac{M_\Phi^2}{q_t^2} \biggr) \\
   &+\frac{3}{2}\cf \ln \biggl( \frac{M_\Phi^2}{q_t^2} \biggr)
    +\cf\z2
                     \biggr)
                                      \biggr].
\end{align}
Keeping with the notation of Ref.~\cite{Field:2004tt}, we write these
coefficients with an overbar as follows

\begin{equation}
\bar{A}^{(1)}_b = \Blue{\cf}, \quad
\bar{B}^{(1)}_b = - \frac{3}{2}\Blue{\cf}, \quad
\bar{C}^{(1)}_{b\bar{b}} = \frac{1}{2}\Blue{\cf}\z2.
\end{equation}
It is important to note that in contrast to $W^\pm/Z^0$ production and 
Drell-Yan processes\cite{Altarelli:1979ub, Arnold:1990yk}, the 
$\bar{C}^{(1)}$ coefficient is positive.

Finally, let us turn to determining the $C^{(1)}$ and $C^{(2)}$ 
coefficients for the total cross-section resummation, although total 
cross-sections will not be presented in this letter. Using the 
results of Ref.~\cite{Harlander:2003ai}, we take the Mellin moments of 
the corrections in the limit $N \rightarrow \infty$ ($z \rightarrow 1$ 
in $z$-space). The NLO corrections are easy to color decompose due to 
the presence of only one color factor\cite{Field:2004tt}. Leaving the 
terms that were originally proportional to the $\delta(1-z)$ factor 
inside curly brackets, we find

\begin{align} \nonumber
\Delta_{b\bar{b}}^{(1)} =
   \biggl[&2 \Blue{\cf} \biggr] \Red{\ln^2(N)}
 + \biggl[ 2 \Blue{\cf} \lnf \biggr] \Red{\ln(N)} \\
 + \biggl\{&2\Blue{\cf}\z2 - \Blue{\cf} + 2 \lnrf \biggr\}
 + 2\Blue{\cf}\z2
\end{align}
where we have given both the renormalization scale $\mu_r$ and the 
factorization scale $\mu_f$ dependence in the results. 

In contrast to NLO, the NNLO corrections contain a mix of color factors
(both $\ca$ and $\cf$ appear). Although it is easy to see that the
factor proportional to $\ln^4(N)$ should clearly be $2\cfcf$, no unique
color decomposition from the results provided in
Ref.~\cite{Harlander:2003ai} can be determined for all the terms in the
expression. However, the numeric result can be written

\small
\begin{align}
\Delta_{b\bar{b}}^{(2)} & = \nonumber
   \biggl[ \frac{32}{9} 
   \biggr] \! \Red{\ln^4(N)}
 + \biggl[ \frac{44}{9} - \frac{8}{27} \nf
         + \frac{64}{9} \lnf
   \biggr] \! \Red{\ln^3(N)} \\ \nonumber
&+ \biggl[ \frac{34}{3} + \frac{92}{9} \z2 - \frac{20}{27} \nf
         + \biggl( \frac{38}{3} - \frac{4}{9} \nf \biggr) \lnr \\ \nonumber
&        - \frac{16}{3} \lnf
         + \frac{32}{9} \lnff
   \biggr] \! \Red{\ln^2(N)}
 + \biggl[ \frac{404}{27} - 14 \z3 \\ \nonumber
&        - \frac{56}{81} \nf
         + \biggl( \frac{34}{3} + \frac{92}{9} \z2
         - \frac{20}{27} \nf \biggr) \lnf 
         - \biggl( 9 - \frac{2}{9} \nf \biggr) \\ \nonumber
&        \times \lnff
         + \biggl( \frac{38}{3} - \frac{4}{9} \nf \biggr) \lnr \lnf
    \biggr] \! \Red{\ln(N)} + \biggl\{ \frac{115}{18} \\ \nonumber
&         + \frac{58}{9} \z2
          - \frac{26}{3} \z3 - \frac{19}{18} \z4
          + \biggl( \frac{2}{27} - \frac{10}{27} \z2
                  + \frac{2}{3} \z3 \biggr) \nf \\ \nonumber
&         + \biggl( \frac{25}{12} + \frac{38}{3} \z2 \biggr) \lnr
          - \biggl( \frac{1}{18} + \frac{4}{9} \z2 \biggr) \nf \lnr \\ \nonumber
&         + \biggl( \frac{11}{12} - 10 \z2 - \frac{122}{9} \z3 \biggr) \lnf
          + \biggl( \frac{1}{18} + \frac{4}{9} \z2 \biggr) \nf \lnf \\ \nonumber
&         + \biggl( \frac{19}{4} - \frac{1}{6} \nf \biggr) \lnrr
          + \biggl( \frac{19}{4} - \frac{32}{9} \z2
                  - \frac{1}{6} \nf \biggr) \lnff \\ \nonumber
&         + \biggl( \frac{1}{3} \nf - \frac{19}{2} \biggr) \lnr \lnf
   \biggr\} 
 + \frac{34}{3} \z2 + \frac{88}{9} \z3 + \frac{364}{45} \zz2 \\ \nonumber
&         - \biggl( \frac{20}{27} \z2 + \frac{16}{27} \z3 \biggr) \nf
          + \biggl( \frac{38}{3} \z2 - \frac{4}{9} \z2 \nf \biggr) \lnr \\
&         - \biggl( \frac{16}{3} \z2 - \frac{128}{9} \z3 \biggr) \lnf
          + \frac{32}{9} \z2 \lnff
\end{align}
\normalsize

We can also determine the NNLL $A^{(2)}$ and $B^{(2)}$ coefficients. 
$A^{(2)}$ agrees with a previous calculation\cite{Balazs:1998sb}. We
find

\begin{align}
A^{(1)}_b &= \Blue{\cf}  \\
A^{(2)}_b &= \frac{1}{2} \Blue{\cf}
             \biggl( 
             \Blue{\ca} \biggl(
                        \frac{67}{18} - \z2
                        \biggr)
             - \frac{10}{9}\Blue{\nf \tr}
             \biggr)
\end{align}
for $A^{(1)}$ and $A^{(2)}$ and

\begin{equation}
B^{(1)}_b = - \frac{3}{2} \Blue{\cf}
\end{equation}

\begin{align} \nonumber
B^{(2)}_b &= \Blue{\cfcf} \biggl( 
                          \frac{3}{2}\z2 - 3 \z3 - \frac{3}{16}
                          \biggr) \\ \nonumber
          &- \Blue{\ca\cf} \biggl(
                          \frac{11}{18}\z2 - \frac{3}{2}\z3 
                                           + \frac{13}{16}
                          \biggr) \\
          &- \Blue{\nf \cf \tr} \biggl( \frac{1}{4} + \frac{2}{9} \z2
                          \biggr)
\end{align}
for the $B^{(1)}$ and $B^{(2)}$ coefficients. The Mellin moments
$\Delta^{(1)}_{b\bar{b}}$ and $\Delta^{(2)}_{b\bar{b}}$ are novel, as is
$B^{(2)}_b$. 

\section{Results and Conclusions}
\label{results}

\begin{figure}
  \begin{center}
    \begin{tabular}{c}
      \resizebox{85mm}{!}{\includegraphics{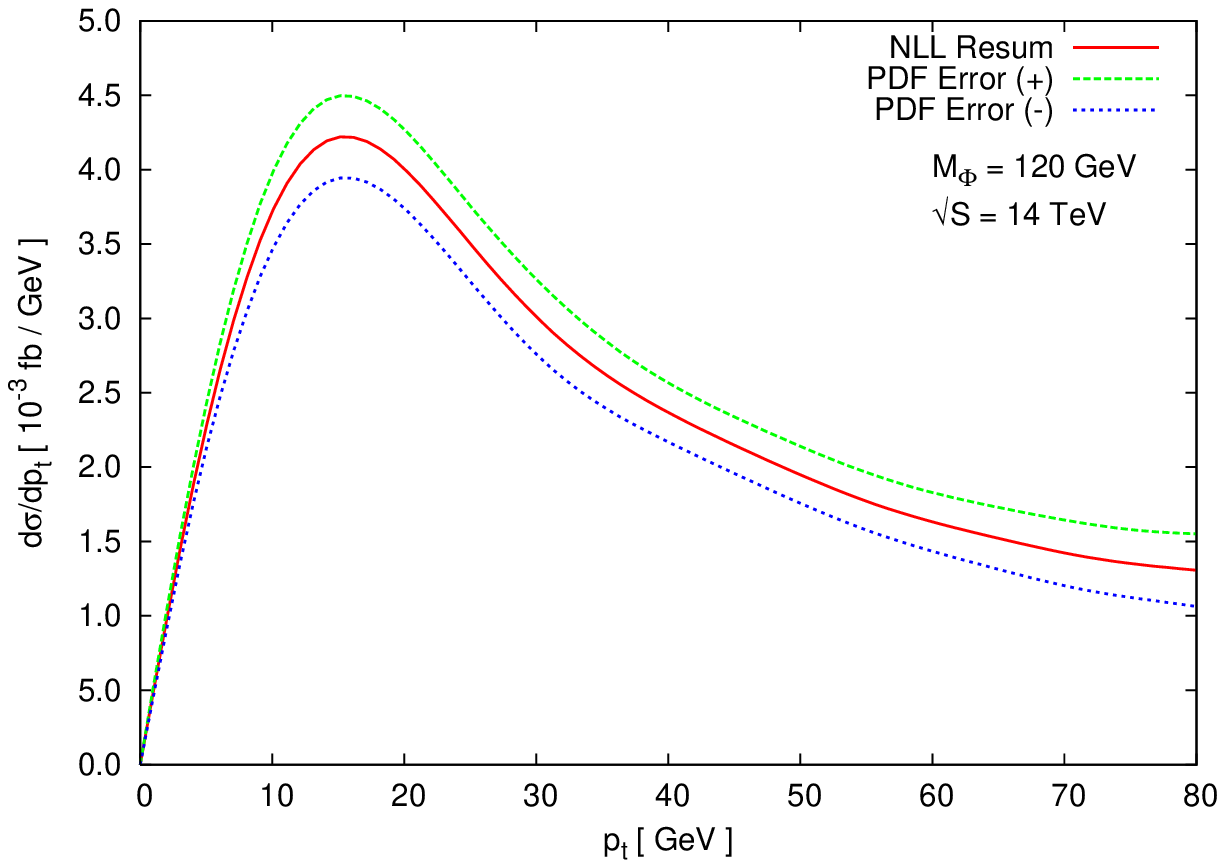}} \\
      (a) \\
      \resizebox{85mm}{!}{\includegraphics{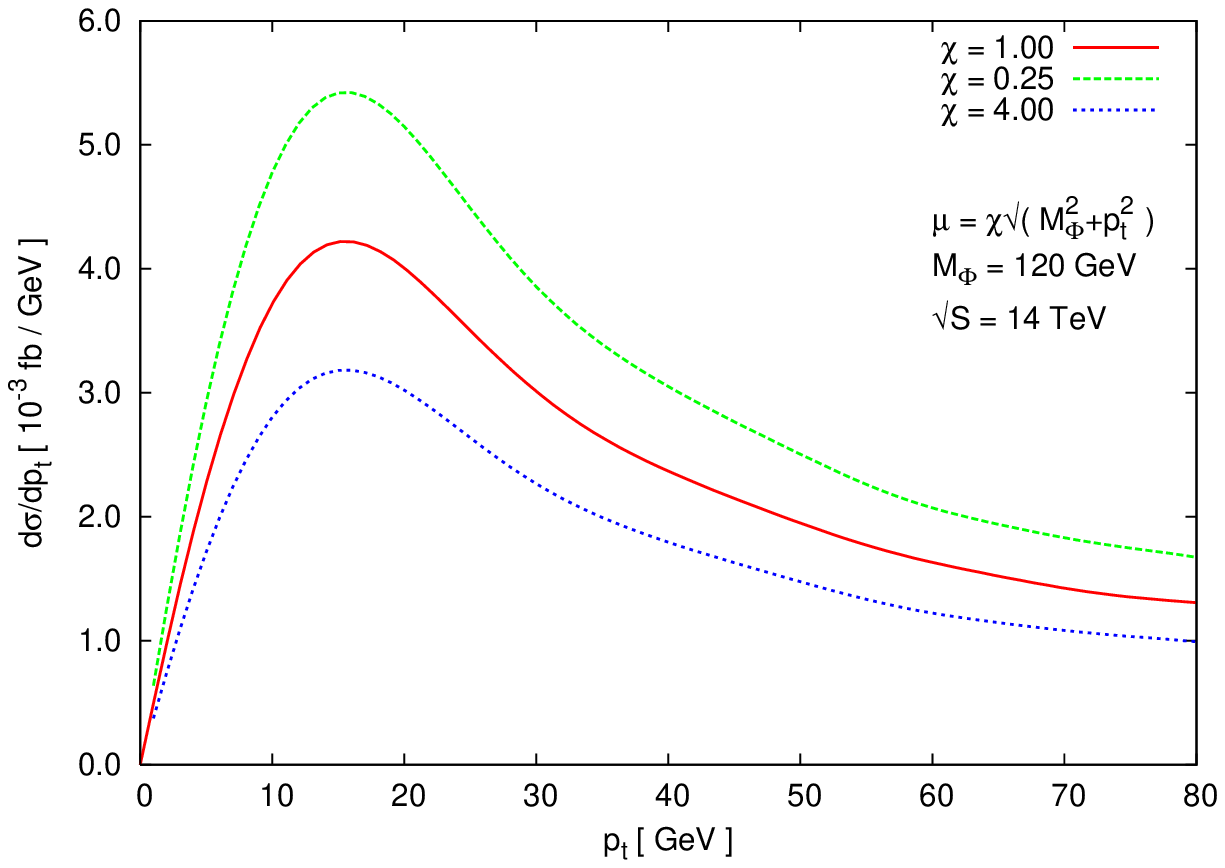}} \\
      (b) \\
      \resizebox{85mm}{!}{\includegraphics{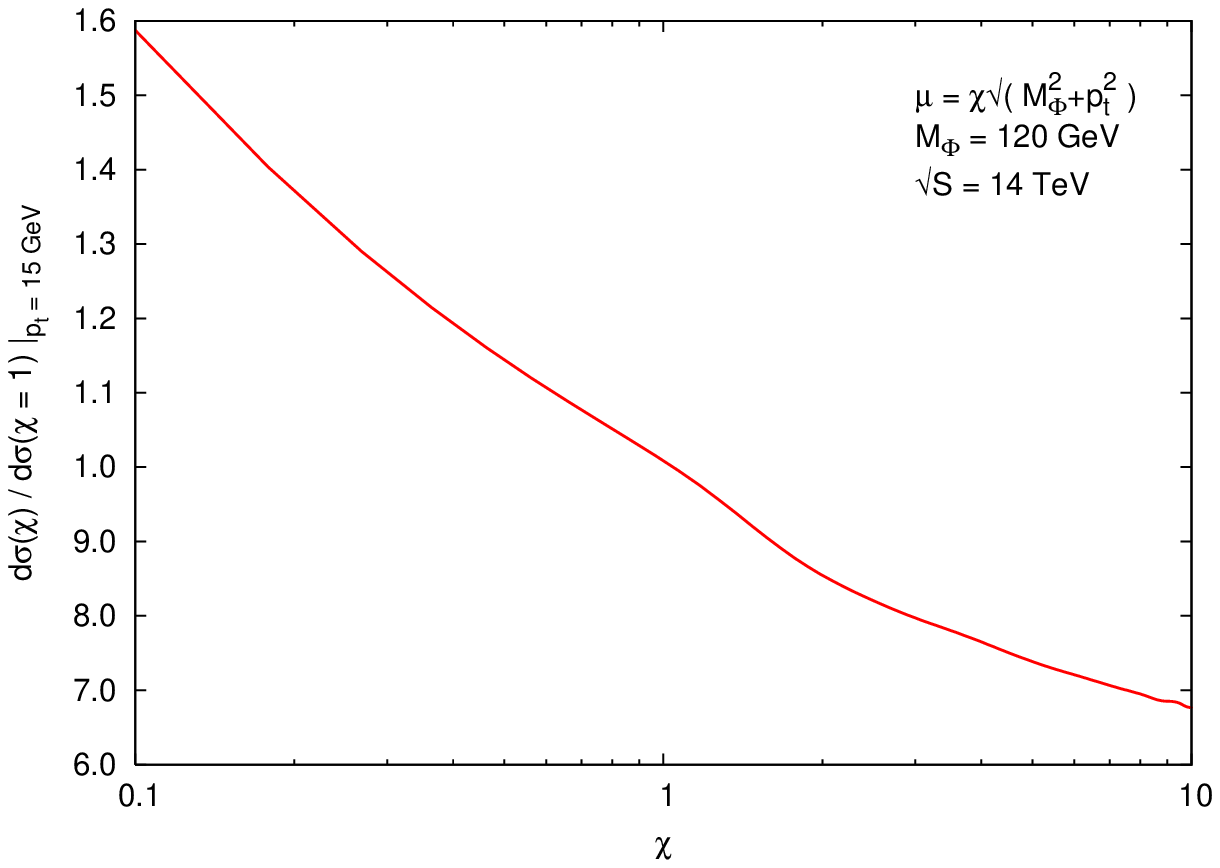}} \\
      (c)
    \end{tabular}
    \caption{Figure~\ref{dsigma}a shows the errors associated with the 
             CTEQ6.1M PDF set. The variation is approximately $8 - 
             12$\%. Figure~\ref{dsigma}b show the variation of the 
             renormalization and factorization scale for a factor of 
             $1/4$ and $4$. These scales were chosen because there has 
             been great interest in the scale $\mu = M_\Phi /4$. We find 
             this variation to be approximately $20$\%.
             Figure~\ref{dsigma}c shows the movement of the peak of the 
             differential distribution (at $15$~GeV) for a variation in 
             the scale by a factor of $10$.}
    \label{dsigma}
  \end{center}
\end{figure}

The differential resummation coefficients and the position of the peak
of the differential cross-section is of great interest to the
experimental community involved with Higgs research at the LHC,
particularly in the $M_\Phi=120$~GeV mass range. Here the Higgs will
decay primarily into $b\bar{b}$ pairs that can be tagged. Knowing where
the peak of the differential distribution lies, especially if it is
below the $p_t$ of a typical trigger event, is of utmost importance.
This letter will help in the analysis of the $b\bar{b}$ initial state.

The results of our calculations can be found in Figure~\ref{dsigma}.  We
have done our analysis for the LHC (a proton-proton collider at
$\sqrt{S}=14$~TeV). We find that the differential distribution at the
LHC peaks at a transverse momentum of approximately $15$~GeV. We find
that the magnitiude of the differential cross-section is an excellent
match with previously published results\cite{Dicus:1998hs,
Balazs:1998sb, Campbell:2002zm}. The results for the Tevatron are
extremely similar, but are smaller by a factor of $60$ and the peak
moves to a transverse momentum of approximately $13$~GeV in the
differential distribution. 

A detailed study of the uncertainties in the calculation show that the
uncertainty due to the PDF set is approximately $8-12$\%. At the peak of
the distribution, the uncertainty is approximately $10$\% due to the
PDFs. When the scale is varied by a factor of four, we see a variation
in the differential cross-section of approximately $20$\%. This would
give us a combined uncertainty of $32$\%, which is slightly better than
the gluon-gluon channel\cite{Field:2004tt} uncertainty in the
differential distribution. However, when the scale is only varied by a
factor of two (as was the case for the gluon-gluon channel), the total
uncertainty drops to $25$\%.

We have calculated the resummation coefficients needed for NLL inclusive
Higgs production via bottom-quark fusion in the \textsc{sm} and the
\textsc{mssm} for the differential cross-section and for the NNLL
resummation for the total cross-section. We find a smaller uncertainty
in the bottom-quark initial state than the gluon-gluon initial state.

\begin{acknowledgments}

The author would like to acknowledge the help and comments of J.~Smith,
S.~Dawson, G.~Sterman, W.~Vogelsong, F.~Olness, and A.~Field-Pollatou. I
would also like to thank W.~Kilgore and R.~Harlander for supplying the
output of their calculation\cite{Harlander:2003ai} including its scale
dependence.

\end{acknowledgments}

\end{document}